\documentclass[12pt]{article}
\usepackage{amssymb}
\begin{document}
\begin{center}

\textbf{CAN THE HIERARCHY PROBLEM BE SOLVED BY 
FINITE-TEMPERATURE  MASSIVE FERMIONS IN THE RANDALL-SUNDRUM MODEL?}

\bigskip

\bigskip

   I. Brevik\footnote{e-mail: iver.h.brevik@mtf.ntnu.no}  \\        

  \bigskip

\bigskip            

          Division of Applied Mechanics,
           Norwegian University of Science and Technology,
           N-7491 Trondheim, Norway \\           
\bigskip
\bigskip

 June 2001
\end{center} 

\bigskip        

\begin{abstract}

    Quantum effects of bulk matter, in the form of massive fermions, are considered in the Randall-Sundrum $AdS_5$ brane world at finite temperatures. The thermodynamic energy (modulus potential) is calculated in the limiting case when the temperature is low, and is shown to possess a minimum, thus suggesting a new dynamical mechanism for stabilizing the brane world. Moreover, these quantum effects may solve the hierarchy scale problem, at quite low temperatures. The present note reviews essentially the fermion-related part of the recent article by I. Brevik, K. A. Milton, S. Nojiri, and S. D. Odintsov, Nucl. Phys. {\bf B 599}, 305 (2001).
\end{abstract}

\bigskip
\bigskip
\bigskip
\bigskip

Submitted to Gravitation and Cosmology (G@C), special issue devoted to Quantum Gravity, Unified Models and Strings, to mark the 100th anniversary of Tomsk State Pedagogical University. Edited by Professor S. D. Odintsov.

\newpage

\section{Introduction}

Consider the Randall-Sundrum (RS) scenario \cite{randall99}, i.e., a non-factorizable geometry with one extra fifth dimension. This dimension, called $y$, is compactified on an orbifold $S^1/Z_2$ of radius $R$ such that $-\pi R \leq y \leq \pi R$. The orbifold fixed points at $y=0$ and $y=\pi R$ are the locations of the two three-branes. We will below, instead of $y$, use the nondimensional coordinate $\phi$, defined by $\phi=y/R$, thus lying in the interval $[-\pi, \pi]$. The RS  5D metric is
\begin{equation}
ds^2=e^{-2kR|\phi|}\eta_{\mu\nu} dx^\mu dx^\nu -R^2 d\phi^2,
\end{equation}
\label{1}
where $\eta_{\mu\nu}$=diag(-1,1,1,1) with $\mu=0,1,2,3$. The 5D metric is $g_{MN}$ with capital subscripts, $M=(\mu, 5)$. The parameter $k \sim 10^{19}$ GeV is of Planck scale, related to the AdS radius of curvature, which is $1/k$. The points $(x^\mu,\phi)$ and $(x^\phi, -\phi)$ are identified. The metric (1) is valid if the 5D cosmological constant $\Lambda$ and the 5D Planck mass $M_5$ are related through 
\begin{equation}
\Lambda=-6M_5^3k^2.
\end{equation}
\label{2}
The cosmological constants at the boundaries have to fulfil the constraints $\Lambda_{0}=-\Lambda_{\pi}=-\Lambda/k$. This is the fine-tuning problem in the RS model. The 4D Planck mass $M_P$ is related to $M_5$ through
\begin{equation}
M_P^2=\frac{M_5^3}{k}\left( 1-e^{-2\pi kR} \right).
\end{equation}
\label{3}
The $\phi=0$ brane (Planck brane) is associated with the mass scale $M_P$, whereas the $\phi=\pi$ brane (TeV-brane) is associated with the scale $M_P e^{-\pi kR}$ lying in the TeV region provided that $kR \simeq 12$.

Assume now that there is a scalar field $\Phi$ in the bulk, with action
\begin{equation}
S_\Phi=\frac{1}{2}\int d^4 x\, \int_{-\pi}^{\pi} d\phi \sqrt{-G}\left\{ G^{AB}\partial_A \Phi \partial_B \Phi
-\left(m^2+\frac{2\alpha k}{R}(\delta(\phi)-\delta(\phi-\pi))\right) \Phi^2 \right\}.
\end{equation}
\label{4}
Here $\alpha$ is a nondimensional constant which parametrizes the mass on the boundaries. All fields in the 5D bulk can be regarded as Kaluza-Klein modes, which in turn can be considered as 4D fields on the brane with an infinite tower of masses. The mass spectrum $m_n$ of the Kaluza-Klein modes in $\Phi$ is given \cite{goldberger00, gherghetta00} by roots of
\begin{equation}
j_{\nu} (x_n)y_{\nu} (ax_n)-j_{\nu} (ax_n)y(x_n)=0.
\end{equation}
\label{5}
Here $x_n=m_n/ak$, $a=e^{-\pi kR}$, $\nu=\sqrt{4+m^2/k^2}$, and the altered Bessel functions are
\begin{equation}
j_\nu(z)=(2-\alpha)J_\nu(z)+zJ_\nu'(z),~~~y_\nu(z)=(2-\alpha)Y_\nu(z)+zY_\nu'(z).
\end{equation}
\label{6}

\section{Effective potential for fermions in 5D AdS space at finite temperature}

Following Ref. \cite{brevik01}, we consider the energy (effective potential) for a bulk quantum field on a 5D AdS background at finite temperature. For a fermion of momentum ${\bf p}$ and mass m the partition function $Z_p^f$ is $2\cosh(\beta E_p/2)$, where $\beta=1/T$ is the inverse temperature and $E_p=\sqrt{{\bf p}^2+m^2}$. The total fermionic partition function $Z^f$ in a three-dimensional volume $V$ then becomes
\begin{equation}
\beta F^f=-\ln Z^f = -V\int \frac{d^3p}{(2\pi)^3}\,\ln \left[ 2\cosh \left( \frac{\beta E_p}{2} \right) \right],
\end{equation}
\label{7}
$F^f$ being the free energy. The corresponding thermodynamic energy $E^f$ is
\begin{equation}
E^f=\frac{\partial}{\partial \beta}(\beta F^f)=-V\int \frac{d^3 p}{(2\pi)^3}\, \frac{E_p}{2} \tanh \left(\frac{\beta E_p}{2} \right).
\end{equation}
\label{8}
As already mentioned, the 5D Kaluza-Klein modes can be considered as 4D fields on the brane with an infinite tower of masses, so by summing up the KK modes given by Eq.~(5) we get the following expression for the total KK energy:
\begin{equation}
E^{fKK}=-{\cal{F}} \left[ \frac{\sqrt{{\bf p}^2+a^2k^2 x^2}}{2}\tanh \left( \frac{\beta}{2}\sqrt{{\bf p}^2+a^2k^2x^2}\right)  \right].
\end{equation}
\label{9}
Here the functional $\cal F$ is defined by
\begin{equation}
{\cal F}[f(p,x)]=V\int \frac{d^3p}{(2\pi)^3}\,\frac{i}{2\pi}\int_C dx\frac{d}{dx}f(p,x)
\ln [j_\nu(x)y_\nu(ax)-j_\nu(ax)y_\nu(x)],
\end{equation}
\label{10}
the contour $C$ encircling the positive real axis.

Since the zero temperature contribution to the energy, $E^f(\infty)$, has been calculated in Ref. \cite{goldberger00a}, we subtract this contribution from $E^f(\beta)$ in Eq.~(8) and consider henceforth
\begin{equation}
\tilde{E}^f(\beta)=E^f(\beta)-E^f(\infty).
\end{equation}
\label{11}
This quantity is finite. In terms of the variable $q$ defined via $|{\bf p}|=q/\beta$ we can write Eq.~(11) as
\begin{equation}
\tilde{E}^f(\beta)=\frac{V}{2\pi^2\beta^4}\int_0^\infty dq\,
 q^2\frac{\sqrt{q^2+\beta^2m^2}}{e^{\sqrt{q^2+\beta^2m^2}}+1}.
\end{equation}
\label{12}
This expression holds for arbitrary temperatures. Assume now that the temperature is low, $\beta \gg 1$. Then it is convenient to change the variable from $q$ to $s$ via $q=\sqrt{s^2+2\beta ms}$, whereby
\begin{eqnarray}
\tilde{E}^f(\beta)&=&\frac{V}{2\pi^2\beta^4}\int_0^\infty ds\,\frac{(s+\beta m)^2\sqrt{s^2+2\beta ms}}{e^{s+\beta m}+1}
                                                   \nonumber \\
                  & &\rightarrow \frac{Vm^{5/2}e^{-\beta m}}{(2\pi \beta)^{3/2}}
\end{eqnarray}
to leading order. We assume that $\beta m \gg 1$; then, because of the factor $e^{-\beta m}$ we need to include only the lowest root $x=x_1$ in Eq.~(5), corresponding to $n=1$. When $a$ is small, this yields $j_\nu(x_1)=0$ (we assume $\alpha +\nu \neq 2$), so that $x_1$ is of order unity. Adding the contribution from the zero-point energy we find the following effective potential:
\begin{equation}
V^f(a)=k^4 B_2^f\left( a^{2\mu}+\frac{B_3^f}{B_2^f}(\beta k)^{-3/2}a^{5/2}e^{-\beta kax_1} \right),
\end{equation}
\label{14}
where $\mu=\nu+2$. We have here defined
\begin{equation}
B_2^f=-\frac{V}{16\pi^2}\frac{2^{1-2\nu}}{\nu \Gamma(\nu)^2}\int_0^\infty dt\,t^{2\nu+3}\,\frac{K_\nu '(t)}{I_\nu '(t)} >0,
\end{equation}
\label{15}
\begin{equation}
B_3^f=\frac{V}{(2\pi)^{3/2}}\,x_1^{5/2}.
\end{equation}
\label{16}
We have now come to the main point: when $\beta k$ is large, the effective potential (14) has a nontrivial {\it minimum}. The order of magnitude of $a$ at the minimum, $a=a_m$, is given roughly by
\begin{equation}
a_m \sim \frac{1}{\beta k} \ln \beta k.
\end{equation}
\label{17}
As mentioned above, taking $kR \simeq 12$ means that the $\phi=\pi$ brane is associated with the TeV region. That means, $e^{-\pi kR}\sim 10^{-17}$. With $k\sim 10^{19}$ GeV we thus see that with $1/\beta \sim 10$ GeV we get $\beta k \sim 10^{18}$, thus $a_m \sim 10^{-17}$ according to Eq.~(17). The weak scale, $a_m k \sim 10^2$ GeV, can in this way be generated.

As a numerical example, assume that $\alpha=2$. If $\nu=5/2$ or $\mu=\nu+2=9/2$ we get $x_1=3.6328$, $B_3^f/B_2^f=11.9408$, and the minimum occurs at $a_m=5.15 \times 10^{-17}$. 

In conclusion, a bulk quantum fermion may generate a thermal (flat) 5D AdS brane-world with the necessary hierarchy scale. This example shows that quantum bulk effects in a brane-world $AdS_5$ at nonzero temperature may not only stabilize the brane-world but also provide the dynamical mechanism for the resolution of the hierarchy problem, with no fine-tuning.

Our example involving a single fermion is somewhat unsatisfactory because the minimum at $a \neq 0$ is only local. A somewhat more extended discussion, involving both bosons and fermions, is given in Ref.~\cite{brevik01}.

We mention that a very recent approach to stabilize the radius of the brane-world $AdS_5$ is the paper of Flachi {\it et al.} \cite{flachi01}. 

Finally, as a general remark, we mention that the RS scenario is of interest also for quantum gravity (for a general treatise on QG, see Ref.~\cite{buchbinder92}).

\end{document}